\documentclass[a4paper,12pt]{article}
\usepackage{amsmath}
\usepackage{graphicx}
\usepackage{amsfonts}
\usepackage{multirow}
\usepackage{cite}

\DeclareFontFamily{U}{rsf}{}
\DeclareFontShape{U}{rsf}{m}{n}{
  <5> <6> rsfs5 <7> <8> <9> rsfs7 <10-> rsfs10}{}
\DeclareMathAlphabet\Scr{U}{rsf}{m}{n}

\mathchardef\varGamma="0100
\mathchardef\varDelta="0101
\mathchardef\varTheta="0102
\mathchardef\varLambda="0103
\mathchardef\varXi="0104
\mathchardef\varPi="0105
\mathchardef\varSigma="0106
\mathchardef\varUpsilon="0107
\mathchardef\varPhi="0108
\mathchardef\varPsi="0109
\mathchardef\varOmega="010A

\newcommand{\be}{\begin{equation}}
\newcommand{\ee}{\end{equation}}
\newcommand{\bea}{\begin{eqnarray}}
\newcommand{\eea}{\end{eqnarray}}

\newcommand{\one}{\mathbb{1}}

\newcommand{\e}{\epsilon}

\newcommand{\epsm}[9]{\ensuremath{\left(\begin{array}{ccc}\e^{#1}&\e^{#2}&\e^{#3}\\\e^{#4}&\e^{#5}&\e^{#6}\\\e^{#7}&\e^{#8}&\e^{#9}\end{array}\right)}}
\renewcommand{\one}{\ensuremath{{\bf 1}}}

\begin{document}
\thispagestyle{empty}
\vspace{-2cm}
\begin{flushright}
{\small
CPHT-RR055.0710\\
LPT-ORSAY 10-58 \\
IFT-7/2010\\
\today}
\end{flushright}
\begin{center}
{\bf \Large  Flavour in supersymmetry: horizontal symmetries  \\
or wave function renormalisation}
\vskip 0.5cm

{\bf
Emilian Dudas$^{+,*}$
Gero von Gersdorff $^{+,}$ ,\\
Jeanne Parmentier $^{+,}$ and Stefan Pokorski $^{\dag}$ \footnote{Hans Fischer Senior Fellow, Institute for Advanced Studies, Technical University, Munich, Germany} \\

}
\vskip 0.5cm
{\it  $^{+}$ Centre de Physique Th\'eorique,
Ecole Polytechnique and CNRS,\\
F-91128 Palaiseau, France \\
$^{*}$ LPT,Bat. 210, Univ. de Paris-Sud, F-91405 Orsay, France \\
$^{\dag}$ Institute of Theoretical Physics, Warsaw University, Hoza 69, 00-681 Warsaw,
Poland  \\
}
\vspace{0.4cm}

{\bf Abstract}
\end{center}
We compare theoretical and experimental predictions of  two main classes of models addressing fermion mass hierarchies and
flavour changing neutral currents (FCNC) effects in supersymmetry: Froggatt-Nielsen (FN) $U(1)$ gauged flavour models and Nelson-Strassler/extra dimensional models with hierarchical wave functions for the families. We show that whereas the two lead to identical predictions in the fermion mass matrices, the second class generates a stronger suppression of FCNC effects. We prove that, whereas at first sight the FN setup is more constrained due to anomaly cancelation conditions, imposing unification of gauge couplings in the second setup generates conditions which precisely match the mixed anomaly constraints in the FN setup. Finally, we provide an economical extra dimensional realisation of the hierarchical wave functions scenario in which the leptonic FCNC can be efficiently suppressed due to the strong coupling (CFT) origin of the electron mass.

\clearpage

\section{Introduction and outline}

The Standard Model (SM)  is successful in describing  the strong suppression of the FCNC and CP violating processes
but this success strongly relies on the pattern of fermion masses and mixing angles taken from experiment. It has since
long been a big theoretical challenge to find extensions of the  SM that address the origin of the Yukawa couplings and
simultaneously solve the hierarchy problem of the SM, in no conflict  with the FCNC and CP violation data.  The flavour
structure of the new physics, needed to explain the pattern of the Yukawa matrices, also has to control the new physics at
TeV scale that protects the Higgs potential from large radiative corrections, so that the  new sources of the
FCNC and CP violation are strongly suppressed.

It is an old and interesting proposal that the flavour dynamics and the hierarchy problem can be simultaneously addressed
in supersymmetric models with spontaneously broken  horizontal gauge symmetries and the Froggatt-Nielsen (FN) mechanism
for Yukawa couplings \cite{fn,ns,ir,binetruy,dps,dps2,Chankowski:2005qp,buras}.  An extensive theoretical and phenomenological
work shows that such models with Abelian  or non-Abelian \cite{nonabelian1,nonabelian2} horizontal symmetries can correctly
reproduce the pattern of Yukawa matrices.  At the same time, they control the flavour structure of the soft supersymmetry
breaking terms in the gravity mediation scenario and can be compatible with very strong experimental constraints from the FCNC and CP
violation sector, without the need to raise the scale of sfermion masses beyond that needed to solve the little hierarchy problem. However, this compatibility often requires restricted range of supersymmetric parameters and/or some additional
structural assumptions  \cite{ns,dps2,kawamura} (see \cite{Lalak:2010bk} for a recent discussion). In general, there is not much room for manoeuvre and one may expect FCNC to be close to the present bounds.

More recently, it has been proposed that the pattern of Yukawa
matrices and the suppression of FCNC
effects in supersymmetric theories can be understood as solely due to
strong wave function renormalisation (WFR models)  of  the matter fields, superimposed on
the initial flavour anarchical structure at the very high (Planck) scale $M_0$  \cite{Nelson:2000sn}.
The origin of such effects could  be RG running down to some scale $M$ few orders of magnitude below $M_0$, with large anomalous
dimensions of the matter fields generated by the coupling of the MSSM
sector to a conformal sector
\cite{Nelson:2000sn,Kobayashi:2001kz,Nelson:2001mq}  or different localisation of different matter fields in a (small) extra dimension introduced just for flavour \cite{Choi:2003fk,nps}. The latter idea has also been extensively discussed as a solution to the flavour problem in non-supersymmetric Randall-Sundrum type models, with strongly warped extra dimensions \cite{csaki}.

As it has already been noticed in the original paper by Nelson and Strassler, the predictions of the WFR mechanism in supersymmetric models resemble those of the flavour models based  on the FN mechanism, with horizontal abelian symmetries, with all SM fermions carrying charges of the same sign and with one familon field.
Indeed, the predictions of the two approaches for the Yukawa matrices are identical, after proper identification of the
corresponding parameters. There is a finite number of FN supersymmetric models (of the horizontal charge assignments) with abelian horizontal symmetry,  one familon field and all fermion charges of the same sign that  are a) theoretically consistent, b) correctly describe  quark and lepton  masses and mixings.  Each horizontal charge assignment can be identified  with concrete values of the set of free parameters in the WFR approach, with the same predictions for the Yukawa matrices. Thus, using the previous results on the FN models we can easily infer a viable set of the WFR models. We point out that this set is likely to be a complete set of such models  if we require gauge coupling unification.

The magnitude of the FCNC and CP violation at the electroweak scale is determined by the coefficients of the dimension 6
operators in the effective SM lagrangian obtained after integrating out the supersymmetric degrees of freedom \cite{masiero}.  As we discuss below in detail, the two approaches differ significantly in their predictions for  the suppression factors for some of the dimension 6 operators.
It is therefore of some interest to compare other predictions, in particular for the FCNC and CP violation suppression,
of the WFR models with predictions of the FN models  that successfully describe fermion masses and mixing. This is the purpose of this note.  It is easy to make such a comparison for  each pair of the models introduced above. However, the
FN models that are successful in the Yukawa sector also include models which have no correspondence to the WFR approach, like models with charges of both signs or models  with several U(1)'s.  Here the comparison is less straightforward but one can see some general differences.

In Sec.~\ref{structure}, putting aside the potential origin of the strong WFR effects that could be responsible for the hierarchy of Yukawa couplings, we compare the structure of the operators violating flavour in the two approaches from a 4d point of view , and discuss their phenomenological predictions. We draw attention to certain important structural differences in favour of WFR, such as
no distinction in wave function renormalisation between fermions and antifermions, no $D$-term contribution to sfermion masses  and no problem with uncontrolled coefficients of order unity.  We notice that whereas significant suppression of FCNC is achieved in the squark sector, the constraints in the leptonic sector coming from $\mu \to e \gamma$ are still difficult to satisfy.

Sec.~\ref{uni} is devoted to the discussion of the gauge coupling unification in the WFR model.  A stunning coincidence is pointed
out with the Green-Schwarz anomaly cancelation conditions in the horizontal symmetry models.

In Sec.~\ref{origin} we discuss in some detail the possible origin of the strong WFR effects. For the extra dimensional interpretation, we comment on the differences and the benefits
compared to RS and propose a CFT origin of the electron mass which has the virtue of decoupling the $A$-term of the electron from its Yukawa coupling. The end result is a strong suppression of the leptonic FCNC violations ($\mu \to e \gamma$) compared to the 4d discussion in Sec.~\ref{structure}.
Sec.~\ref{conclusions} contains our conclusions.
\section{Horizontal symmetry versus WFR:  structure and predictions}
\label{structure}

We consider effective supersymmetric models with softly broken supersymmetry, described by a K\"ahler potential and a superpotential, below the flavour symmetry breaking scale $M$ but above the soft supersymmetry breaking scale $M_{susy}$.
The flavour structure may be present in the kinetic terms, the superpotential (in general, non-renormalisable) and in the pattern of soft terms.
We concentrate only on models with positive FN charges, which are relevant for the WFR case.
The effective action is determined by
\begin{eqnarray}
&& W = \epsilon^{q_i+u_j+h_u} (Y^U_{ij} + A^U_{ij} X ) Q_i  U_j H_u + \epsilon^{q_i+d_j+h_d} (Y^D_{ij} + A^D_{ij} X ) Q_i  D_j H_d \nonumber \\
&&+  \epsilon^{l_i+e_j+h_d} (Y^E_{ij}  + A^E_{ij} X ) L_i E_j H_d \nonumber \\
&& K = \epsilon^{|q_i-q_j|} (1 + C_{ij} X^{\dagger} X ) Q_i^{\dagger} Q_j + \cdots \ , \label{h1}
\end{eqnarray}
where $\epsilon = \theta/M$, with $\theta$ a chiral (super)field of $U(1)$ charge $-1$, $X = \theta^2 F$ is the SUSY breaking spurion and all flavour matrix elements $Y^U_{ij}$, etc are considered to be of order one.
The family charges of fermion superfields are  defined as
$q_i$ for the flavour components of the left-handed doublet $Q_L$, and $u_i$ and $d_i$ for the
flavour components of the (left-handed) quark singlet fields $U^c$ and $D^c$, the charge conjugate of the right-handed flavour triplets $U_R$ and $D_R$, respectively, and similarly for leptons.
Horizontal charges are defined in some electroweak basis. In that basis, flavour mixing is present also in the kinetic terms.
However, the rotation to the canonical basis does not change the leading powers of $\epsilon$ in the rest of the lagrangian
(we assume all coefficients $C_{ij}$, $Y_{ij}$ and $A_{ij}$ to be of $\mathcal O(1)$)  and we shall always refer to the canonical basis.

In the WFR case  the effective action at the scale   $M$ is determined by
\begin{eqnarray}
&& W = (Y^U_{ij} + A^U_{ij} X ) Q_i  U_j H_u + (Y^D_{ij} + A^D_{ij} X ) Q_i  D_j H_d  \nonumber \\
&&+ (Y^E_{ij}  + A^E_{ij} X )  L_i E_j H_d \nonumber \\
&& K = \epsilon^{- 2 q_i} Q_i^{\dagger} Q_i +  C_{ij} X^{\dagger} X  Q_i^{\dagger} Q_j    \cdots  \ . \label{h2}
\end{eqnarray}
Here the factors $\epsilon^{- 2 q_i}$   are  the wave function renormalisation factors, originating from the physics
between $M_0$ and $M$, and in the notation suitable for the comparison of the two approaches. Any order unity flavour
mixing in the kinetic terms at the scale $M_0$ has already been rotated away.
After rescaling of the wave functions $Q_i \to \epsilon^{q_i} Q_i$, etc (also including the possiblity of rescaling of the Higgs fields),  the effective action in the WFR case is given by
\begin{eqnarray}
&& W = \epsilon^{q_i+u_j+h_u} (Y^U_{ij} + A^U_{ij} X ) Q_i  U_j H_u +  \epsilon^{q_i+d_j+h_d} (Y^D_{ij} + A^D_{ij} X ) Q_i  D_j H_d  \nonumber \\
&& + \epsilon^{l_i+e_j+h_d} (Y^E_{ij} + A^E_{ij} X )  L_i  E_j H_d \nonumber \\
&& K = Q_i^{\dagger} Q_i +  C_{ij} \epsilon^{q_i+q_j} X^{\dagger} X  Q_i^{\dagger} Q_j    \cdots  \ . \label{h3}
\end{eqnarray}

The comparison of the two approaches is immediate.  For the two models to give identical predictions for Yukawa couplings at the high scale the parameters of the supersymmetric WFR models are fixed in terms of the charge assignment in the FN models.  However, since the wave function renormalisation does not distinguish between particles and antiparticles,
the suppression of sfermion masses is much stronger in the WFR case. Similar observation in the non-SUSY case  has been made in \cite{davidson}.
Actually the class of FN models which really compare directly to WFR models are the ones with only one $U(1)_X$, positive charges and with only one familon field of negative charge breaking it, such that all Yukawas are generated by holomorphic couplings to the familon.

For any comparison with experimental data we have to be in the basis where the quark mass matrices are diagonal. Since the main experimental constraints come from the down quark sector it is very convenient to remain in an electroweak basis (for explicit $SU(2)\times U(1)$ gauge invariance) but the one in which the down quark Yukawa matrix is diagonal. Thus, the scalar field terms in the Lagrangian are subject to (appropriate) left and right rotations that diagonalise down quark Yukawa matrix.  Such rotations, acting on the off-diagonal terms in the sfermion mass matrices
do not change their leading suppression factors (powers of $\epsilon$)  but generically are the source of additional contributions to the off-diagonal terms coming from the splitting in the diagonal entries. For FN models, the two obvious sources of
the diagonal splitting are potentially  flavour dependent order unity coefficients $C_{ii}\neq C_{jj}$ and generically present  flavour dependent U(1) $D$-terms.  Those additional  contributions to the off-diagonal terms are of the order  of the rotation angles diagonalising
the down quark Yukawa matrix (roughly speaking of the order of the CKM angles)  and are unwelcome.
They provide an upper bound (in fact  uncomfortably strong) on the suppression of the off-diagonal terms. It so happens that in the discussed here models with all fermions carrying the same sign horizontal charges the  suppression factors  of the original  off-diagonal terms are the same as the suppression of the terms originating from the diagonal splitting
so the problem of compatibility with the data is similarly difficult for both components (see \cite{Lalak:2010bk}).
However, there are $U(1)$ models with  charges of both signs and/or  with several $U(1)$ symmetries which
do not have WFR counterpart but are often successful in the Yukawa sector and give strong suppression of
the original flavour off-diagonal  sfermion mass terms. Still, they face the mentioned above generic problem of the flavour dependent $D$-term contribution to the diagonal masses  and possible diagonal splitting by uncontrolled by the $U(1)$ symmetry
coefficients of order unity. After rotations, the suppression of the off-diagonal terms in the quark mass eigenstate basis is then similar as in the models with the same sign charges
and results in certain tensions in the parameter space of the soft supersymmetry breaking terms \cite{Lalak:2010bk}.
It is clear that WFR approach avoids those problems. There is no $U(1)$ symmetry and no $D$-terms and the flavour dependent diagonal terms in the sfermion mass matrices
are suppressed by powers of $\epsilon$, so there is also no problem of uncontrolled coefficients of order unity.

In addition to working in the electroweak basis with diagonal down quarks, for a meaningful comparison with experimental data we have to include all the MSSM-like renormalisation effects for the running from the scale $M$ to the electroweak scale. Finally, the standard analysis of the FCNC and CP violation data  is performed in terms of the coefficients of the dimension 6
operators in the effective SM lagrangian obtained after integrating out the supersymmetric degrees of freedom \cite{masiero}.  The coefficients of those operators are calculable in terms of the  soft supersymmetry breaking parameters
and the discussed above suppression factors have a direct correspondence in the suppression factors of the  higher dimension operators.

Let us compare the flavour properties of some models of fermion mass hierarchies under the two paradigms of family symmetry and WFR. From our discussion above it is clear that once we have a FN model with all fermion charges of the same sign that  correctly reproduces the fermion mass hierarchy it can immediately be translated into a WFR model.\footnote{FN are seemingly more constrained, as some form of anomaly cancelation has to be imposed. We will see in Sec.~\ref{uni} that preservation of the successful MSSM gauge coupling unification in WFR models places very similar constraints on the assignment of "charges"(suppression factors) in the latter case.}
In the following we shall compare some of the corresponding pairs of models mentioned above. As we discussed earlier,
for a global picture one should also compare the set of  viable WFR models with FN models that do not have any WFR
correspondence but are successful in the Yukawa sector, too.  However, after inclusion  of the effects of the splitting on the diagonal and of the $D$-term contributions to the sfermion masses such models give predictions for FCNC very close to the
same sign charge models, so we don't discuss them any more. From the point of view of proton decay operators, both approaches can
generate some suppression: $U(1)_X$ FN can also completely kill proton decay if for example the lepton charges are
$l_i= n_i+ 1/3, e_i = m_i-1/3$, with $n_i,m_i$ integers (all other MSSM charges being integers), since then there is an effective
$Z_3$ discrete leptonic symmetry protecting the proton to decay. More generally, both FN and WFR  generate some suppression for the first generations due to their large charges.

The flavour suppression is parameterised by the variable $\epsilon$ introduced earlier.
We have in mind $\epsilon$ to be of the order the Cabbibo angle, $\e\sim 0.22$, but certainly other values can be considered provided one appropriately rescales the charges.
Consistent charge assignments have for instance been classified in Refs.~\cite{dps,Chankowski:2005qp}. Here we will consider 3 models:\footnote{A and B are taken from Ref.~\cite{Chankowski:2005qp}, where they are called models 1 and 5 respectively, model C was studied in Ref.~\cite{dps}.}
\begin{gather}
q=u=e=(3,2,0)\,,\quad d=\ell=(2,0,0)+d_3\,, \tag{Model A}\\
q=u=e=(4,2,0)\,,\quad d=\ell=(1,0,0)+d_3\,,\tag{Model B}\\
q=(3,2,0)\,,\quad u=(5,2,0)\,,\quad d=(1,0,0)+d_3\,,\quad \ell= q + \ell_3\,,\quad e= d-\ell_3 \,.\tag{Model C}
\end{gather}
In all three cases the horizontal charges of the Higgs fields are zero.
Notice that the choice $q_3=u_3=0$ is a requirement for obtaining a heavy top, while the freedom in $d_3$ is related to $\tan\beta $ via the bottom Yukawa:
\be
\e^{-d_3}\,\tan\beta \sim\frac{m_t(M_c)}{m_b(M_c)}\sim \e^{-3}\,.
\label{tanb}
\ee
The resulting Yukawa couplings for model A are displayed in the last column of Tab.~\ref{tabA}. They readily reproduce the observed masses and mixings of the SM fermions for suitable choices of $O(1)$ coefficients.

\newcommand{\topline}{\hline\\[-.2cm]}
\newcommand{\sep}{\\[-.2cm]\hline\\[-.2cm]}
\newcommand{\bottomline}{\\[-.2cm]\\\hline}

\begin{table}
\begin{center}
\begin{tabular}{c|ccc}
\topline
$a$&$\tilde m^2_{a,LL}\,/\,m_0^2$&$\tilde m^2_{a,RR}\,/\,m_0^2$
&$ A_{a}\,/\,m_0\sim Y_a $\\
\sep
&$r_q\,\one +\epsm{6}{5}{3}{5}{4}{2}{3}{2}{0}$
&$r_u\,\one +\epsm{6}{5}{3}{5}{4}{2}{3}{2}{0}$
\\[-.8cm]
$u$&&&$\epsm{6}{5}{3}{5}{4}{2}{3}{2}{0}$\\[-.8cm]
&
$r_q\,\one +\epsm{0}{1}{3}{1}{0}{2}{3}{2}{0}$
&$r_u\,\one +\epsm{0}{1}{3}{1}{0}{2}{3}{2}{0}$
\\
\sep
&$r_q\,\one +\epsm{6}{5}{3}{5}{4}{2}{3}{2}{0}$
&$r_d\,\one +t_\beta^2\epsm{10}{8}{8}{8}{6}{6}{8}{6}{6}$
\\[-.8cm]
$d$&&&$t_\beta \epsm{8}{6}{6}{7}{5}{5}{5}{3}{3}$\\[-.8cm]
&$r_q\,\one +\epsm{0}{1}{3}{1}{0}{2}{3}{2}{0}$
&$r_d\,\one +\epsm{0}{2}{2}{2}{0}{0}{2}{0}{0}$
\\
\sep
&$r_\ell\,\one +t_\beta^2\epsm{10}{8}{8}{8}{6}{6}{8}{6}{6}$
&$r_e\,\one +\epsm{6}{5}{3}{5}{4}{2}{3}{2}{0}$\\[-.8cm]
$e$&&&$t_\beta \epsm{8}{7}{5}{6}{5}{3}{6}{5}{3}$\\[-.8cm]
&$r_\ell\,\one +\epsm{0}{2}{2}{2}{0}{0}{2}{0}{0}$
&$r_e\,\one +\epsm{0}{1}{3}{1}{0}{2}{3}{2}{0}$
\bottomline
\end{tabular}
\caption{Yukawas and soft scalar mass squared matrices for model A \cite{Chankowski:2005qp}: $q=u=e=(3,2,0)$, $\ell=d$ with $d-d_3=(2,0,0)$ and we have used the relation $\tan\beta=\e^{d_3-3}$.   The upper row corresponds to a WFR model, while the lower one to a FN one.}
\label{tabA}
\end{center}
\end{table}

After inclusion of the renormalisation effects \cite{Louis:1995sp}  , the soft mass terms at the electroweak scale are
to a good approximation given by
\bea
\tilde m_{u, LL, ij}^{2}&\sim&  r_q\, m_{1/2}^2\,\delta_{ij}+\hat m_q^2 \epsilon^{|q_i\pm q_j|}\\
\tilde m_{u, RR, ij}^{2}&\sim&  r_u\, m_{1/2}^2\,\delta_{ij}+\hat m_u^2 \epsilon^{|u_i\pm u_j|}\\
\tilde m_{u, LR, ij}^{2} &\sim& A_u v\sin\beta\,\epsilon^{q_i+u_j}
\eea
\bea
\tilde m_{d, LL, ij}^{2}&\sim&  r_q\, m_{1/2}^2\,\delta_{ij}+\hat m_q^2 \epsilon^{|q_i\pm q_j|}\\
\tilde m_{d, RR, ij}^{2}&\sim&  r_d\, m_{1/2}^2\,\delta_{ij}+\hat m_d^2 \epsilon^{|d_i\pm d_j|}\\
\tilde m_{d, LR, ij}^{2} &\sim& A_d v\cos\beta\,\epsilon^{q_i+d_j}
\eea
\bea
\tilde m_{e, LL, ij}^{2}&\sim&  r_\ell\, m_{1/2}^2\,\delta_{ij}+\hat m_\ell^2 \epsilon^{|\ell_i\pm\ell_j|}\\
\tilde m_{e, RR, ij}^{2}&\sim&  r_e\, m_{1/2}^2\,\delta_{ij}+\hat m_e^2 \epsilon^{|e_i\pm e_j|}\\
\tilde m_{e, LR, ij}^{2} &\sim& A_e v\cos\beta\,\epsilon^{\ell_i+e_j}
\eea
where we have defined the high scale soft masses $m_{1/2}$, $\hat m_a$ and $A_a$. Baring some additional suppression mechanism, a completely natural theory would require all these terms to be of the same order, and we will henceforth set them all equal to a common mass $m_0$.
The terms that are suppressed by powers of $\e$ are multiplied by flavour dependent $\mathcal O(1)$ coefficients that are omitted here for clarity. The charges are all positive or zero, and the positive sign applies to the WFR case whereas the negative one to the FN case.
The constants $r_a$ parameterize the leading gauge renormalization and are given approximately by $r_q=6.5$, $r_u=6.2$, $r_d=6.1$, $r_\ell=0.5$ and $r_e=0.15$. Yukawa corrections are expected to be important for the third generation but given the unknown $\mathcal O(1)$ coefficients of the tree level soft mass matrices they are irrelevant for our discussion. The resulting soft mass matrices are also displayed in Tab.~\ref{tabA}.
Several points deserve to be emphasized.
\begin{itemize}
\item
For WFR all 1st and 2nd generation mass eigenvalues (and, in fact, some 3rd generation ones as well) are predicted from the running of gauge/gaugino loops, while in the FN case  the explicit tree level soft masses  give non-negligible contribution, particularly to slepton masses.
\item
Yukawas and chirality changing soft masses ($A$-terms) receive the same suppression, and are in fact the same for both FN and WFR.
\item
In the LL and RR sectors the off-diagonal masses are more suppressed for WFR than for FN, as explained above.
\end{itemize}

To compare with experiment, bounds are usually given for the mass insertion parameters $\delta^a_{ij}$ at a reference sfermion mass. They are defined as
\be
\delta^a_{MN,ij}=\frac{\tilde m^{2}_{a,MN,ij}}{\tilde m_{a,M,i}\,\tilde m_{a,N,j}}\,,\qquad
\langle\delta^a_{ij}\rangle=\sqrt{\delta^a_{LL,ij}\delta^a_{RR,ij}}
\ee
for $a=u,d,e$, $M,N=L,R$, $i\neq j$. The expressions are normalized to  the diagonal entries $\tilde m_{a,M,i}$. To the $A$ terms one associates analogous parameters (for any $i,j$):
\be
\delta^a_{LR,ij}=(\delta^a_{RL,ji})^*=\frac{\tilde m^2_{a,LR,ij}}{\tilde m_{a,L,i}\,\tilde m_{a,R,j}}\,.
\ee

Starting with the limits from the hadron sector, we give the bounds and the results for Model A in Tab.~\ref{hadLLA} and \ref{hadLRA}. All bounds in Tab.~\ref{hadLLA} are comfortably satisfied (even for large $\tan\beta$) and in fact would allow for a much smaller squark mass. 
Notice that in the FN model with analogous charge assignments it is very difficult to satify the bound on $\langle\delta_{12}\rangle$ \cite{Lalak:2010bk}.
Since the squark mass mixing between the first two generations is suppressed at most by two powers of $\epsilon$, to satisfy
the bound one needs very strong flavour blind renormalisation effects, i.e.~a large ratio of of the initial values of the gluino mass to the squark mass at the very high scale.

The chirality flipping mass insertions of Tab~\ref{hadLRA} are more constraining. In particular, the 11 entries are strongly constrained from the EDM measurements of the neutron. Nevertheless, the corresponding predictions of our model for 1 TeV squark mass marginally satisfy the experimental bounds.

Turning to leptons, we quote in Tab.~\ref{lepA} the bounds~\footnote{Note that the decay rate depends on the sum $(\delta_{LR,ij})^2+(\delta_{RL,ij})^2$ \cite{masiero}. The $LL$ and $RR$ entries are much less constrained and we will not consider them here.} resulting from LFV decays of the charged leptons, $\mu\to e\gamma$, $\tau\to e\gamma$ and $\tau\to \mu\gamma$  and the theoretical predictions obtained under the assumption of a universal supersymmetry breaking scale $m_0$ at high energy.  Then, at the electroweak scale  $A_e\sim m_0$ and the typical scale for sleptons is $\tilde m_{sl}=(r_\ell r_e)^{\frac{1}{4}}m_0$.
One observes that  even for the slepton mass as high as $400$ GeV (corresponding the $m_0=750$ GeV) the contribution to $\mu\to e\gamma$ is not sufficiently suppressed.  It is interesting to know how far one can  adjust the charges $e_i$ and $\ell_i$ to ameliorate this problem. To this end, consider the product
\be
\delta^e_{LR,12}\delta^e_{RL,12}\sim \frac{A_e^2v^2}{\tilde m_{sl}^4}\e^{\ell_1+e_2+\ell_2+e_1}\cos^2\beta
\sim
\frac{A_e^2m_e m_\mu}{\tilde m_{sl}^4}\,.
\label{one}
\ee
It is therefore clear that this product is independent of the concrete
charge assignment
and can only be lowered by increasing $\tilde m_{sl}$ or decreasing $A_e$.
This means that at least one of the individual contributions is bigger than
\be
\frac{A_e\sqrt{m_e m_\mu}}{\tilde m_{sl}^2}\sim 3.5\times 10^{-5}
\label{two}
\ee
where the numbers are for $\tilde m_{sl}=400$ GeV. This is a robust prediction (up to $\mathcal O(1)$ coefficients), and indeed Tab.~\ref{lepA} shows that it holds in particular for model $A$.
A stronger suppression  can be obtained only if $A$ terms are  smaller than $m_0$ and/or  $m_0$ has larger value, i.e.
$\tilde m_{sl}>400$ GeV.
For instance one can get an acceptable decay rate for $A_e\sim 100$ GeV and $\tilde m_{sl}\sim 400$ GeV.
Lowering the slepton mass further requires more and more fine tuning of $A_0$, while $\tilde m_{sl}\sim 400$ GeV implies squark masses of the order of 1.9 TeV which is uncomfortably large for the little hierarchy problem.
In conclusion, the leptonic bounds are more constraining than the hadronic ones ( see also \cite{Nelson:2000sn}, \cite{nps}).
Finally, let us stress that FN models do possess a similar problem (with identical bounds on the $LR/RL$ sector). In addition, they predict insufficient suppression in the $LL$ and $RR$ sectors.

In Sec.~\ref{origin} we will point out a novel mechanism to suppress  the $\mu\to e\gamma$ decay rate, opening the possibility to lower the superpartner masses without fine-tuning the leptonic $A$ terms  in the WFR models.

\begin{table}
\begin{center}
\begin{tabular}{cc|cc|cc|cc}
\hline
$a$	&$ij$		&\multicolumn{2}{|c|}{$\delta^a_{LL,ij}$}	
				&\multicolumn{2}{|c|}{$\delta^a_{RR,ij}$}
				&\multicolumn{2}{c}{$\langle \delta^a_{ij}\rangle$}
				\\
&& Exp.& Th &Exp.&Th.&Exp.&Th.\\
\hline
$d$	&12		&0.03	&$8.6\times 10^{-5}$	&0.03	&$9.1\times 10^{-7}\, t_\beta^2
			$&0.002	&$8.9\times 10^{-6}\, t_\beta$\\
$d$	&13		&0.2	&$1.8\times 10^{-3}$	&0.2	&$9.1\times 10^{-7}\, t_\beta^2$	
			&0.07	&$4.0\times 10^{-6}\, t_\beta$\\
$d$	&23		&0.6	&$8.1\times 10^{-3}$	&1.8	&$1.8\times 10^{-5}\, t_\beta^2$	
			&0.2	&$3.8\times 10^{-4}\, t_\beta$\\
\hline
$u$	&12		&0.1	&$8.6\times 10^{-5}$	&0.1	&$8.6\times 10^{-5}\, $	
			&0.008	&$8.6\times 10^{-5}$\\
\hline
\end{tabular}
\caption{Bounds on hadronic chirality-preserving mass insertions and results from WFR with model A. Bounds (taken from Tab.~IV of Ref.~\cite{Isidori:2010kg}) are valid for a squark mass of 1 TeV and scale linearly with the latter.}
\label{hadLLA}
\end{center}
\end{table}

\begin{table}
\begin{center}
\begin{tabular}{cc|cc|cc}
\hline
$a$	&$ij$		&\multicolumn{2}{|c|}{$\delta^a_{LR,ij}$}	
				&\multicolumn{2}{|c}{$\delta^a_{RL,ij}$}
				\\
&& Exp.& Th. &Exp.&Th.\\
\hline
$d$	&12		&$2\times10^{-4}$	&$8.1\times 10^{-6}$	&$2\times 10^{-3}$	&$1.8\times 10^{-6}\,$\\
$d$	&13		&0.08	&$8.1\times 10^{-6}$	&0.08	&$3.7\times 10^{-5}\,$\\
$d$	&23		&0.01	&$3.7\times 10^{-5}$	&0.01	&$7.6\times 10^{-4}\,$\\
$d$ &11		&$4.7\times 10^{-6}$	&$3.9\times 10^{-7}$	
			&$4.7\times 10^{-6}$	&$3.9\times 10^{-7}$\\
\hline
$u$	&12		&0.02	&$3.7\times 10^{-5}$	&0.02	&$3.7\times 10^{-5}\, $	\\
$u$	&11		&$9.3\times 10^{-6}$	&$8.1\times 10^{-6}$	
			&$9.3\times 10^{-6}$	&$8.1\times 10^{-6}$\\
\hline
\end{tabular}
\caption{Bounds on hadronic chirality-flipping mass insertions and results from WFR with model A. Bounds taken from Tab.~V of Ref.~\cite{Isidori:2010kg} are valid for a squark mass of 1 TeV. While the bounds on the $i\neq j$ ($i= j$) elements grow linearly (quadratically) with the latter, our predictions go down linearly.}
\label{hadLRA}
\end{center}
\end{table}

\begin{table}
\begin{center}
\begin{tabular}{c|ccc}
\hline
$ij$			&\multicolumn{3}{|c}{$\delta^e_{MN,ij}$}
				\\
		&Exp.	&Th. (LR)	&Th. (RL)\\
\hline
12	&$4.8\times 10^{-6}$	&$2.0\times 10^{-5}$	&$9.4\times 10^{-5}$\\
13	&$1.8\times 10^{-2}$	&$4.3\times 10^{-4}$	&$9.4\times 10^{-5}$\\
23	&$1.2\times 10^{-2}$	&$8.9\times 10^{-3}$	&$4.3\times 10^{-4}$\\\hline
\end{tabular}
\caption{Experimental bounds on leptonic mass insertions and results from WFR with model A. Bounds (taken from Tab.~7 of Ref.~\cite{masiero}, using updated bounds on the branching ratios \cite{Amsler:2008zzb}) are valid for a slepton mass of 400 GeV.}
\label{lepA}
\end{center}
\end{table}

\section{Unification and wave-function hierarchies}
\label{uni}

The physical gauge coupling in a supersymmetric field theory is given by \cite{sv,kl}
\begin{equation}
\frac{4 \pi^2}{g_a^2 (\mu)} \ = \ Re f_a \ + \ \frac{b_a}{4} \log \frac{\Lambda^2}{\mu^2} +
\frac{T (G_a)}{2} \log g_a^{-2} (\mu^2) - \sum_r \frac{T_a (r)}{2} \log \det Z_{(r)} (\mu^2) \ , \label{kl1}
\end{equation}
where
\begin{equation}
b_a \ = \ \sum_r n_r T_a (r) - 3 T (G_a) \quad , \quad T_a (r) \ = \ Tr_r \ T_{(a)}^2 \  \label{kl2}
\end{equation}
are the beta function and the Dynkin index of the representation $r$ under the gauge group factor $G_a$, $f_a$ are the
holomorphic gauge couplings,
$Z_{(r),ij}$
are wave functions of matter fields of flavour indices $i,j$
and the determinant $\det Z_{(r)} (\mu^2)$ is taken in the flavour space. \\
   In our case $Z_{(r)} \simeq diag \ (\epsilon^{- 2 q_1^{(r)}} , \epsilon^{- 2 q_2^{(r)}}, \epsilon^{- 2 q_3^{(r)}})$ and therefore
\begin{equation}
   \log \det Z_{(r)}  \ = \ - 2 \sum_i q_i^{(r)} \log \epsilon \ , \label{kl3}
\end{equation}
where $q_i^{(r)}$ are the " $U(1)$ charges" of the matter representations $r = Q,U,D,L,E,H_u,H_d$. Let us define in what follows
the quantities
\begin{equation}
A_a \ = \ - \frac{1}{\log \epsilon} \sum_r \frac{T_a (r)}{2} \log \det Z_{(r)} \ , \label{kl4}
\end{equation}
which are proportional to the additional contribution to the running coming from a strongly coupled sector, producing the hierarchical
wave functions. Notice that usual MSSM unification is preserved if
\begin{equation}
A_3 \ = \ A_2 \ = \ \frac{3}{5} A_1 \ . \label{kl5}
\end{equation}
With the field content of MSSM, we find
\begin{eqnarray}
&& SU(3) \ : \qquad A_3 \ = \ \sum_i (2 q_i + u_i + d_i) \ , \nonumber \\
&& SU(2) \ : \qquad A_2 \ = \ \sum_i (3 q_i + l_i) + h_u + h_d \ , \nonumber \\
&& U(1)_Y \ : \qquad A_1 \ = \ \sum_i (\frac{1}{3} q_i + \frac{8}{3} u_i + \frac{2}{3} d_i + l_i + 2 e_i) + h_u + h_d \ . \label{kl6}
\end{eqnarray}
Notice also that the quantities $A_i$ can be related simply to the determinants of the mass matrices of the quarks and leptons via
\begin{eqnarray}
&& \det (Y_U Y_D^{-2} Y_L^3) \ = \ \epsilon^{\frac{3}{2} (A_1 + A_2 - 2 A_3)} \ , \nonumber \\
&& \det (Y_U Y_D) \ = \ \epsilon^{A_3 + 3 (h_u+h_d)} \ . \label{kl7}
\end{eqnarray}
The reader familiar with the gauged Froggatt-Nielsen $U(1)$ generating Yukawa hierarchies probably recognized already
(\ref{kl5})-(\ref{kl7}), see \cite{ir}, \cite{binetruy}, \cite{dps}. It is worth pointing out the interesting analogy with our present case:
\begin{itemize}
\item In the gauged FN case the quantities (\ref{kl6}) are  precisely the coefficients of the $U(1)_X G_a^2$ mixed  anomalies,
between the gauged $U(1)_X$ and the SM gauge group factors $G_a = SU(3), SU(2), U(1)_Y$.
\item In the FN case (\ref{kl5}) represent the universal (for the heterotic strings) Green-Schwarz anomaly cancelation conditions.
\end{itemize}
In our case,  (\ref{kl5}) represent the unification conditions for the gauge couplings at the energy scale where the strong sector
decouples from the running. Interestingly enough, even if there is no gauge $U(1)$ symmetry in our case, unification of gauge
couplings requires the "charges" determining the wave function  renormalisation to satisfy exactly the same constraints as anomaly cancellation for the U(1) charges in the gauged FN case ! \\
By using the results of \cite{binetruy}, \cite{dps} on the structure of quark and lepton masses, one useful relation can be written
\begin{equation}
A_1 + A_2 - \frac{8}{3} A_3 \simeq 2 (h_u+ h_d) \ . \label{kl8}
\end{equation}
The unification conditions (\ref{kl5}) lead therefore to $h_u+ h_d=0$. Since in the WFR models all "charges" are positive or zero, this
means that $h_u=h_d=0$.  Therefore in the  extra dimensional interpretation of the WFR models (to be discussed in the next section) both Higgs doublets are localized on the UV brane. \\
 Let us notice that, in the FN case, the mixed anomaly conditions (\ref{kl6}) imposed to the model C of Section (\ref{structure}) gives the result $h_u+h_d=0$, $d_3-l_3= 2/3$. A simple solution is $h_u=h_d=0$, $d_3=1,l_3= 1/3$. In this case, the $U(1)_X$ symmetry
 breaks to a discrete $Z_3^L$ acting on the leptons, which protects proton decay. \\
\section{Extra dimensional model for the WFR}
\label{origin}

There are various possible origins for the WFR: 4d strongly coupled or higher-dimensional with flavour-dependent wave-function localization. We use here a variant of the RS setup \cite{rs1}, with an UV brane with energy scale $\Lambda_{UV} $ and an
IR brane with energy scale $\Lambda_{IR} \sim M_{GUT}$. The fifth dimension is therefore very small and the hierarchy is given by
\be
\e=\frac{\Lambda_{IR}}{\Lambda_{UV}}=e^{-k\pi R}
\ee
All MSSM fields live in the bulk \cite{tony}.
Following \cite{Choi:2003fk}, start with K\"ahler terms ($0<y<\pi R$)
\begin{multline}
\hat K=
e^{(1-2c_{h_u})k y }H_u^\dagger H_u+
e^{(1-2c_{h_d})k y }H_d^\dagger H_d\\
+e^{(1-2c_{q,i})k y }Q^\dagger_iQ_i+
e^{(1-2c_{u_i})k y }U^\dagger_iU_i+
e^{(1-2c_{d_i})k y }D^\dagger_iD_i\\
+\delta(y)k^{-3}X^\dagger X\left(
  C_{q,ij}Q^\dagger_iQ_i
+ C_{u,ij}U^\dagger_iU_i
+ C_{d,ij}D^\dagger_iD_i
+ C_{h_u}H_u^\dagger H_u+C_{h_d}H_d^\dagger H_d\right)
\,.
\label{kahler}
\end{multline}
where $i,j=1,2,3$ running over families and the coefficients are flavour anarchic $\mathcal O(1)$ numbers. We have kept only fields with zero modes, the conjugate fields $\phi^c$ have Dirichlet boundary conditions $(--)$ and hence have no zero modes.  The leptons have an analogous Lagrangian. Brane localized kinetic terms can also be introduced, even with arbitrary flavour dependence, without changing the outcome.
We will introduce a superpotential
\begin{multline}
\hat W=\delta(y) k^{-\frac{3}{2}}\left( \hat Y^u_{ij}H_u Q_i U_j+\hat Y^d_{ij}H_d Q_i D_j+k^{-1}X \hat A^u_{ij} H_u Q_iU_j
+ k^{-1}X \hat A^d_{ij} H_d Q_iD_j \right)\\
+\delta(y-\pi R) (k\e)^{-\frac{3}{2}}\left( \hat Y'^u_{ij}\e^{c_{h_u}+c_{q_i}+c_{u_j}}H_u Q_i U_j+\hat Y'^d_{ij}\e^{c_{h_d}+c_{q_i}+c_{d_j}}H_d Q_i D_j \right)
\label{super}
\end{multline}

Notice that we have confined the SUSY breaking spurion $X$ to the UV brane at $y=0$. We have introduced arbitrary dimensionless Yukawa couplings on both branes. After integrating over the extra dimension, the kinetic terms pick up wave function renormalisations
\be
Z_{q}=\frac{1}{(1-2c_q)k}\left(
\e^{2c_q-1}-1
\right)\,,
\\
\ee
and therefore
\be
Z_q \sim \frac{\epsilon^{2c_q-1}}{(1-2c_q)k} \ \ {\rm for } \ c < 1/2 \quad {\rm and} \quad Z_q \sim \frac{1}{(2c_q-1)k} \ \ {\rm for} \
c > 1/2 \ .
\ee
Notice that
\begin{itemize}
\item
For $c_q<\frac{1}{2}$ the field is localized near the IR brane.
 We assign it "charges" $q=\frac{1}{2}-c_q>0$ and $q'=0$.
\item
For $c_q>\frac{1}{2}$ the field is localized near the UV brane.
The charges are $q=0$ and $q'=c_q-\frac{1}{2}>0$.
\item
Exact UV (IR) brane localization is obtained by formally sending $q'$ ($q$) to infinity.
\end{itemize}
After switching to canonical normalization, this leads to Yukawa couplings
\bea
Y_{ij}^u&=& \hat Y^u_{ij}\e^{q_i+u_j+h_u}
+\hat Y'^u_{ij}\e^{q'_i+u'_j+h'_u}\,,\label{yuku}\\
Y_{ij}^d&=&
 \hat Y^d_{ij}\e^{q_i+d_j+h_d}
+\hat Y'^d_{ij}\e^{q'_i+d'_j+h'_d}\label{yukd}\,.
\eea
Each field either suppresses $\hat Y$ or $\hat Y'$, depending on whether it is UV or IR localized.

Since we will take the $X$ field localized on the UV brane, the physical soft masses and $A$ terms at the high scale are given by the expressions in Sec.~\ref{structure} with
\bea
m_0\sim \frac{|F_X|}{k}\,.
\eea

We consider the following localisation of the MSSM fields\footnote{Similar localization of the MSSM flavours from a different perspective
was also considered recently in \cite{tony}.} :
\begin{itemize}
\item
 the first two generations of quarks and leptons are localized near the IR brane. In a holographic 4d interpretation, the first
two generations are composite states.
\item the top quark is localized on or near the UV brane, whereas bottom and tau are localized near the UV brane or near the IR brane, depending
on $\tan \beta$. In the  holographic language, the top quark is therefore elementary.
\item the two Higgs doublets $H_u,H_d$ are localized near the UV brane and therefore have $h_u,\ h_d=0$. They are then elementary from the 4d holographic point of view. In the scenario below, we will consider a finite $h_d'$ describing a non-negligible "tail" near the IR.
\item
the spurion $X$ is located on the UV brane
\end{itemize}
One important point to mention here is that  the extra dimensional realisation of the WFR approach allows for certain generalisations. They are equivalent (and the analogy with FN is true) only  if all Yukawas  are  localized on the UV brane
that is if we neglect the corrections coming from the "tail" of the  Higgs fields near the IR brane.

By comparing with the standard RS non-SUSY setup with fermion mass hierarchies generated by wave functions overlap, we notice
that in the standard RS case, since $\epsilon_{RS} = \frac{\Lambda_{UV}}{\Lambda_{IR}} \sim 10^{-16}$, the bulk masses $c_i$ have to be
tuned close to $1/2$ in order not to generate too big hierarchies in the fermion masses. In our case, we choose to work with a very small extra dimension $10^{-3 }\le \epsilon \le 10^{-1}$ and therefore there is no need for such a tuning. Of course, such a small warping
does not provide a solution to the hierarchy problem anymore, but since we have low-energy supersymmetry, the strong warping is clearly not needed.

Provided $h_u'$ and $h_d'$ are large enough (sharp UV localization), the Yukawa couplings originating from the IR brane (i.e.~the terms proportional to $\hat Y'$ in Eqns.~(\ref{yuku}) and (\ref{yukd})) are always subleading compared to  the ones from the UV brane and hence irrelevant. For moderately large values they can become comparable\footnote{This "switching behavior" was exploited in Ref.~\cite{Agashe:2008fe} to generate an anarchical neutrino spectrum and large mixing angles.}, at least for the light generations, and can in fact be exploited to circumvent the $\mu\to e\gamma$ problem pointed out at the end of Sec.~\ref{structure}. For instance, for all 3 generations of leptons IR-localized (small to moderate $\tan\beta$), one has
\be
Y^e_{ij}= \hat Y^e_{ij}\e^{\ell_i+e_j}+\hat Y'^e_{ij}\e^{h'_d}\,,
\qquad
A^e_{ij}= m_0\hat A^e_{ij}\e^{\ell_i+e_j}
\label{YA}
\ee
Ideally, we would like to suppress the dangerous $A$ terms without suppressing the corresponding Yukawas. This is easy to do: Let us imagine that we increase $\ell_1$ and/or $e_1$ such that $A^e_{12}$ and $A^e_{21}$ are sufficiently small in order to satisfy the bounds for a given slepton mass.\footnote{As a bonus, the $A^e_{11}$ term, responsible for generating an electron EDM, receives additional suppression.} Of course, this will result in a too small electron mass unless we impose that $h_d'$ is responsible for generating $Y^e_{11}$ from the IR brane. We thus choose charges such that
\bea
\ell_1+e_1&>& h_d'\\
h_d'& \sim & 5+\ell_3+e_3\\
\ell_2+e_2& \sim & 2+\ell_3+e_3\,.
\eea
where the last two relations ensure the correct $e-\tau$ and $\mu-\tau$ mass ratios. A possible choice, satisfying also the unification conditions Eq.~(\ref{kl5}), reads
\begin{gather}
q=(4,2,0)\,,\qquad u=(3,2,0)\,,\qquad e=(5,2,0)\,,\\
d=(5,0,0)+d_3\,,\qquad \ell=(4,0,0)+d_3\,,\qquad h_d'\sim 5+d_3\,,
\end{gather}
leading to Yukawas
\be
Y^u\sim\epsm{7}{6}{0}{5}{4}{2}{3}{2}{0},\quad
Y^d\sim t_\beta \epsm{\underline 8}{7}{7}{\underline 8}{5}{5}{8}{3}{3},\quad
Y^e\sim t_\beta \epsm{\underline 8}{\underline 8}{7}{8}{5}{3}{8}{5}{3}\,.
\ee
The underlined exponents are the ones generated from the new contributions in the second term in Eq.~(\ref{YA}). One sees that only the down and the electron masses are affected by $h_d'$. On the other hand, the A terms are given by
\be
A^u\sim Y^u\,,\quad
A^d\sim t_\beta \epsm{12}{7}{7}{10}{5}{5}{8}{3}{3}\,,\quad
A^e\sim t_\beta \epsm{12}{9}{7}{8}{5}{3}{8}{5}{3}\,.
\ee
The suppression in the $12$ and $21$ elements of $A^e$ is now sufficient for a slepton mass around 200 GeV. Notice that the FN models, even with multiple $U(1)$'s, have no analogue of this mechanism.

Notice that in order to forbid R-parity violating operators we need to impose R-parity as symmetry of the effective action. Once this is done, there are usually still dangerous dimension five operators. In our case, if the triplet Higgs fields are localized on the UV brane along with the doublets, these operators are naturally generated there, and we find
\begin{equation}
\frac{1}{\Lambda_{UV}} \epsilon^{q_i + q_j + q_l + l_m} Q_i Q_j Q_k L_m \quad , \quad   \frac{1}{\Lambda_{UV}}
\epsilon^{u_i + u_j + d_k + e_m} U_i U_j D_k E_m  \ . \label{xtra2}
\end{equation}
Due to the localization of the first two generations on the IR brane, we get an additional suppression, as for the UV localized Yukawas
of the first two generations, which is enough in order to bring these operators into their experimental bounds \cite{barbier}.

Finally, extra dimensional interpretation may shed some light on the stunning coincidence of the anomaly cancelation
conditions and the conditions for the gauge coupling unification discussed in the previous section.
The charges $q_i = 1/2-c_i$ can be understood, in a holographic 4d picture, as dimensions of CFT operators that the bulk fields couple to. So it seems that the analog of the gauged $U(1)_X$ is actually the SUSY partner of the dilatation current, the R-symmetry current $U(1)_R$. On the other hand, the anomalies $U(1)_R G_a^2$ are indeed related to the beta functions and therefore to the running of the gauge couplings of the SM gauge factors $G_a$.

\section{Conclusions}
\label{conclusions}

Supersymmetric models with WFR reproduce the success of the FN models for fermion masses with mixings, alleviating at the same time their FCNC
problems. Whereas from a 4d perspective, the improvement in the quark sector is phenomenologically quite successful, in the leptonic
sector there are still problems with $\mu \to e \gamma$. We however showed that in an extra dimensional realization similar to RS but
with an IR brane of mass scale of the order of $M_{GUT}$, with the first two generations composite (IR localized)  and the third one
elementary (UV localized), the problem can be elegantly solved by generating the electron mass by the strong coupling in the CFT sector.
Indeed, the A-term for the electron is strongly suppressed since the supersymmetry breaking spurion field is elementary and the
corresponding terms in the action (as well as the other soft breaking terms) are localized on the UV brane. More generally, the analogy
between FN and WFR is precise in the warped 5d realization when all Yukawa couplings are elementary (UV localized), whereas strong
coupling contributions (IR CFT contributions) add new structure compared to the FN setup.
As a side comment, we notice that similarly, we can generate a $\mu$-term on the IR boundary with large suppression factor if $h'_u,h'_d \gg 0$ in the sense discussed in Section \ref{origin}. This is of course useful  only if for some reason such a term is absent on the UV brane.

We showed that whereas the FN gauge $U(1)$ case is constrained by the various gauge anomaly cancelation conditions, in the WFR case most of these conditions re-emerged in Section \ref{uni} as conditions for gauge coupling unification.
More precisely, the same conditions for the corresponding parameters as the mixed anomaly conditions $A_3 \sim U(1)_X SU(3)^2$, $A_2 \sim U(1)_X SU(2)^2$, $A_1 \sim U(1)_X U(1)_Y^2$ for the U(1) charges appear
in the threshold corrections to the gauge couplings (\ref{kl1}), (\ref{kl4}). They are then constrained by the unification of the SM gauge couplings precisely in the same way as the U(1) charges by the universal Green-Schwarz anomaly cancelation conditions in the FN case.
The mixed anomaly $U(1)_X^2 U(1)_Y$ does not emerge in the WFR setup however and it is therefore still true that in the WFR case
the "charges" $q_i \leftrightarrow c_i $ are less constrained than $U(1)_X$ charges in the FN setup.

One should also mention that in the FN case $U(1)_X$ can be broken to discrete symmetries $Z_N$ which can have nice features like suppressing proton decay at acceptable levels. There does not seem to be analog of this phenomenon in the WFR case.

On the side of the  phenomenological predictions of the WFR scheme and the possibility of its experimental verification,  one sees that FCNC effect are much more strongly suppressed than in the FN models. Thus,  contrary to the predictions of the FN models, one does not expect the FCNC effects to be close to the present bounds  (perhaps with exception of the muon decay). However, there is an interesting correlation between the supersymmetric models for flavour
and the pattern of superpartner masses.  The WFR scheme predicts  all superpartner masses, except the stop masses,
in terms of the gluino mass. In particular, also slepton masses are predicted in terms of the gluino mass.

Finally, it would be interesting to investigate the issue of flavor violation in F-theory models, where similarly there is an analog
of the WFR of generating Yukawa hierarchies \cite{vafa} and a different, gauged FN setup generating them \cite{palti}. 
\subsection*{Acknowledgments}

We thank Tony Gherghetta and Claudio Scrucca for stimulating and helpful discussions.
The work presented was supported in part by the European ERC Advanced Grant 226371 MassTeV, by the CNRS PICS no. 3059 and 4172,
by the grants ANR-05-BLAN-0079-02, the PITN contract PITN-GA-2009-237920, the IFCPAR CEFIPRA programme 4104-2
and by the MNiSZW grant N N202 103838 (2010-2012).
SP thanks the Institute for Advanced Studies at TUM, Munich, for its support and hospitality.


\end{document}